\RequirePackage{fix-cm}
\documentclass[smallextended]{svjour3}       % onecolumn (second format)
\smartqed  % flush right qed marks, e.g. at end of proof
\usepackage{graphicx}

\usepackage{subfig}
\usepackage{makecell}
\usepackage[nolist]{acronym}

\usepackage[%
	square,        % for square brackets
	comma,         % use commas as separators
	numbers,       % for numerical citations;
	sort&compress, % as sort but in addition multiple numerical citations
]{natbib}
\usepackage{float}
\usepackage{url}
\usepackage{multirow}

\usepackage{xcolor}
\usepackage{siunitx}
\usepackage[dash,dot]{dashundergaps}

\newcommand{\figref}[1]{Fig.~\ref{#1}}
\newcommand{\secref}[1]{Section~\ref{#1}}
\newcommand{\tabref}[1]{Table~\ref{#1}}

\usepackage{contour}
\usepackage[normalem]{ulem}

\contourlength{0.8pt}

\renewcommand{\underline}[1]{%
	\uline{\phantom{#1}}%
	\llap{\contour{white}{#1}}%
}

\begin{document}

\begin{acronym}[]
	\acro{AAAR}{Abdominal Aortic Aneurysm Repair}
	\acro{AAA}{Abdominal Aortic Aneurysm}
	\acro{AC}{Alternating Current}
	\acro{ANN}{Artificial Neural Network}
	\acro{CMM}{Coordinate Measuring Machine}
	\acro{DC}{Direct Current}
	\acro{DOF}{Degree of Freedom}
	\acroplural{DOF}{Degrees of Freedom}
	\acro{DNN}{Deep Neural Network}
	\acroplural{DNNs}{Deep Neural Networks}
	\acro{EKF}{Extended Kalman Filter}
	\acro{EMT}{Electromagnetic Tracking}
	\acro{EMTS}{Electromagnetic Tracking System}
	\acro{EVAR}{Endovascular Aneurysm Repair}
	\acro{FFNN}{Feed Forward Neural Network}
	\acro{FPGA}{Field Programmable Gate Array}
	\acro{HP-MIS}{High Precision Minimally Invasive Surgery}
	\acro{LOS}{Line of Sight}
	\acro{MC}{Monte Carlo}
	\acro{MIS}{Minimally Invasive Surgery}
	\acro{MSE}{Mean-Squared-Error}
	\acro{OR}{Operating Room}
	\acro{ROI}{Region of Interest}
	\acro{SLAM}{Simultaneous Localization and Mapping}
\end{acronym}

\title{Leveraging Spatial Uncertainty for Online Error Compensation in EMT
}

%\titlerunning{Understanding Spatial Uncertainty of Error Compensation in EMT}        % if too long for running head

\author{
	Henry Krumb \and
	Sofie Hofmann \and
	David Kügler \and
	Ahmed Ghazy \and
	Bernhard Dorweiler \and
	%Judith Bredemann \and
	Robert Schmitt \and
	Georgios Sakas \and
	Anirban Mukhopadhyay
}


\institute{%
	H. Krumb \and S. Hofmann \and G. Sakas \and A. Mukhopadhyay \at Department of Computer Science, Technische Universit\"at Darmstadt, Germany\\
	\email{henry.john.krumb@gris.tu-darmstadt.de}
	\and%
	D. K\"ugler \at DZNE Bonn, Germany
	\and%
	A. Ghazy \at Klinik für Herz-, Thorax- und Gef\"a{\ss}chirurgie, Universit\"atsmedizin Mainz, Germany
	\and
	B. Dorweiler \at Uniklinik K\"oln
	\and%
	%J. Bredemann \and
	R. Schmitt \at WZL Aachen, RWTH Aachen, Germany
}

\date{Received: date / Accepted: date}

\maketitle

\begin{abstract}
% Purpose
\textbf{\\* Purpose:}
\ac{EMT} can potentially complement fluoroscopic navigation, reducing radiation exposure in a hybrid setting.
Due to the susceptibility to external distortions, systematic error in \ac{EMT} needs to be compensated algorithmically.
Compensation algorithms for \ac{EMT} in guidewire procedures are only practical in an online setting.
%\smallskip

% Methods
\textbf{\\* Methods:}
We collect positional data and train a symmetric \ac{ANN} architecture for compensating navigation error.
The results are evaluated in both online and offline scenarios and are compared to polynomial fits.
We assess spatial uncertainty of the compensation proposed by the \ac{ANN}.
Simulations based on real data show how this uncertainty measure can be utilized to improve accuracy and limit radiation exposure in hybrid navigation.
%\smallskip

% Results
\textbf{\\* Results:}
\acp{ANN} compensate unseen distortions by more than 70\%, outperforming polynomial regression.
Working on known distortions, \acp{ANN} outperform polynomials as well.
We empirically demonstrate a linear relationship between tracking accuracy and model uncertainty.
The effectiveness of hybrid tracking is shown in a simulation experiment.
%\smallskip

% Conclusion
\textbf{\\* Conclusion:}
\acp{ANN} are suitable for \ac{EMT} error compensation and can generalize across unseen distortions.
Model uncertainty needs to be assessed when spatial error compensation algorithms are developed, so that training data collection can be optimized.
Finally, we find that error compensation in \ac{EMT} reduces the need for x-ray images in hybrid navigation.

%\smallskip

%Insert your abstract here. Include keywords, PACS and mathematical
%subject classification numbers as needed.
\keywords{Electromagnetic Tracking \and Hybrid Navigation \and Metallic Distortion Compensation \and Uncertainty Analysis}

\acresetall % we need to re-define after abstract

\end{abstract}

\section{Introduction}
% !TeX spellcheck = en_US

%%%%% background %%%%%
\ac{EMT} is a key technology to enable navigation in minimally invasive surgery without line of sight.
As miniaturized sensors can be integrated into catheters, \ac{EMT} has potential to be employed for guidewire navigation in \ac{AAAR} \cite{manstad2011, manstad2012}.
In current clinical practice, fluoroscopic x-ray imaging is considered the gold standard for guidewire navigation in endovascular aneurysm repair \cite{dijkstra2011intraoperative}.
However, x-ray imaging exposes both the surgeon and the patient to ionizing radiation \cite{giordano2008cervical}.
The high accuracy \cite{kugler2018i3posnet} and visual feedback of fluoroscopy means complete removal of x-ray in minimally invasive vascular surgery is unrealistic in near future.
A more realistic approach is to consider a \textit{hybrid} navigation framework.
In this framework, continuous navigation will be performed by radiation-free \ac{EMT} while x-ray snapshots will be acquired on demand for recalibration or dexterous maneuver.
This hybrid navigation reduces the amount of x-ray images that need to be captured during the procedure, which in turn will reduce the radiation exposure for both surgeon and patient.

EMT navigation is negatively affected by the presence of metal or electromagnetic interference within the vicinity of the tracking system.
The presence of the c-arm x-ray unit within the \ac{OR} is a dominant source of metallic distortion for the EMT measurement.
While it is well known that passive countermeasures (such as removal of the metallic object) might mitigate such error \cite{franz2014}, the c-arm fluoroscopy unit is essential for hybrid tracking procedures.
Thus, the c-arm cannot be removed from the \ac{OR}.
\newcommand{\tabitem}{~~\llap{\textbullet}~~}
\begin{table}[b]
	\begin{tabular}{l|l|l}
		Type of error & Sources & Countermeasures \\
		\hline
		\hline
		\shortstack[l]{system inherent\\errors} & \shortstack[l]{noise,\\fabrication inaccuracies} & \shortstack[l]{averaging, filtering,\\system design improvement} \\
		\hline
		\shortstack[l]{field distortion\\errors} & \shortstack[l]{ferromagnetic/conductive materials,\\electric currents} & \shortstack[l]{active \cite{kindratenko2000survey} or passive \cite{reichl2013}\\compensation}  \\
		\hline
		\shortstack[l]{errors during\\data acquisition} & \shortstack[l]{operator error,\\ phantom uncertainty} & \shortstack[l]{data validation,\\phantom calibration\cite{kuegler2019}} \\
		\hline
		\shortstack[l]{compensation-\\inherent errors} & \shortstack[l]{lack of training data,\\ sparsity of training points} & \shortstack[l]{more training data points\\denser spacing of points}
	\end{tabular}
	\caption{%
		Types of error in the active \ac{EMT} compensation pipeline.
	}
	\label{tab:countermeasures}
\end{table}

Rather, an active error compensation is necessary to improve electromagnetic tracking accuracy.
Unlike random error that can be eliminated by averaging recorded sensor data over multiple samples, compensating systematic error requires more sophisticated algorithms.
Classical techniques such as lookup-tables \cite{raab1979}, interpolation \cite{zachmann1997} or polynomial regression \cite{ikits2001} only work under known distortion characteristics.
Such \textit{offline} compensation requires a tedious data acquisition procedure every time the c-arm position is changed.
Clearly, these algorithms are impractical for hybrid navigation in the \ac{OR}.
\ac{EMT} navigation in surgery thus requires \textit{online} compensation approaches, where the compensating algorithm can be used in any distortion scenario.
Training data for online compensation need to be collected only once for several scenarios.
For the sake of brevity, all types of errors related to \ac{EMT} and countermeasures are summarized in Table \ref{tab:countermeasures}.

\begin{figure}[t]
	\centering
	\begin{minipage}{0.45\linewidth}
		\centering
		\subfloat[\label{fig:phantom}]{\includegraphics[height=0.7\linewidth]{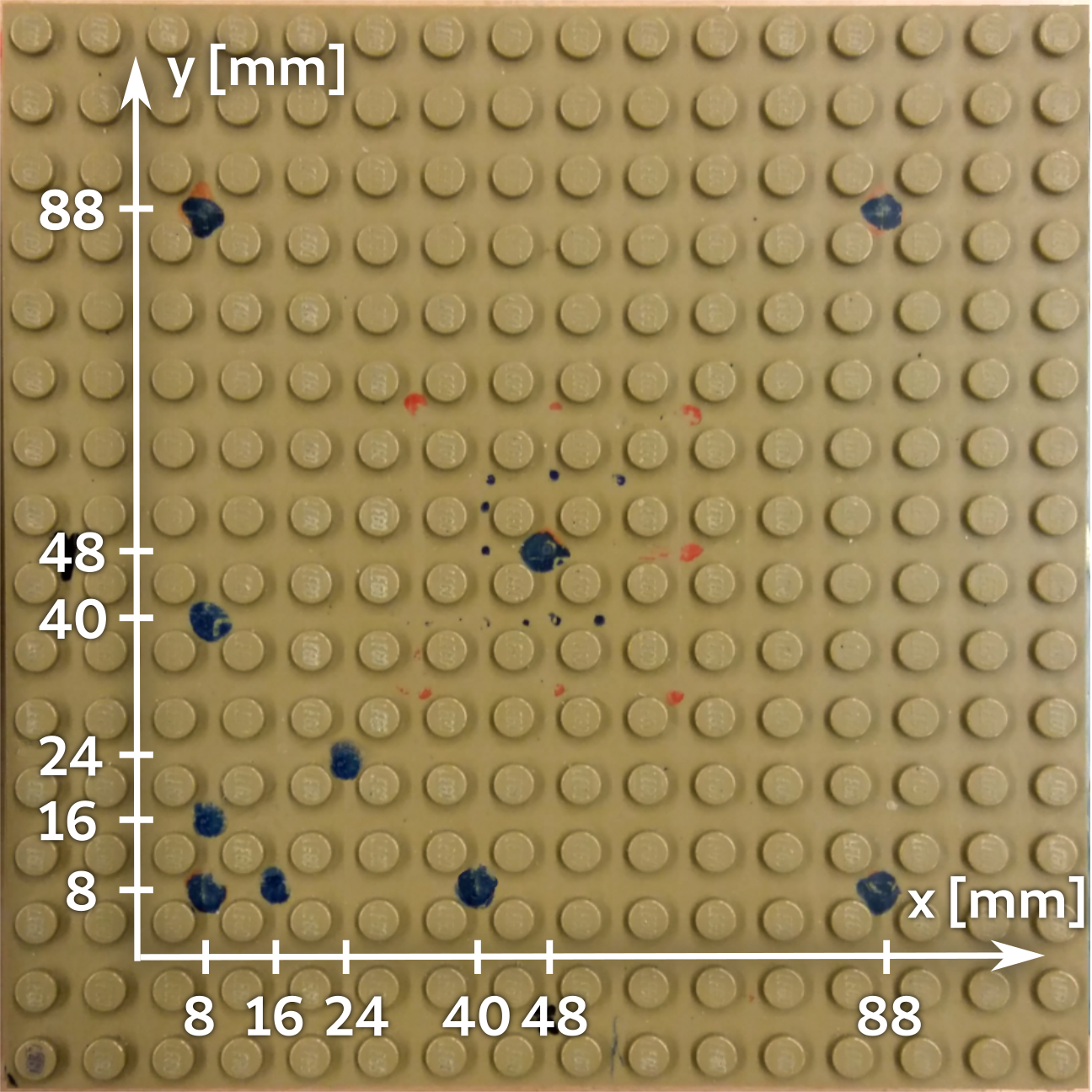}}\vspace{-0.0005\linewidth}
		\subfloat[\label{fig:carm_near}]{\includegraphics[width=0.7\linewidth]{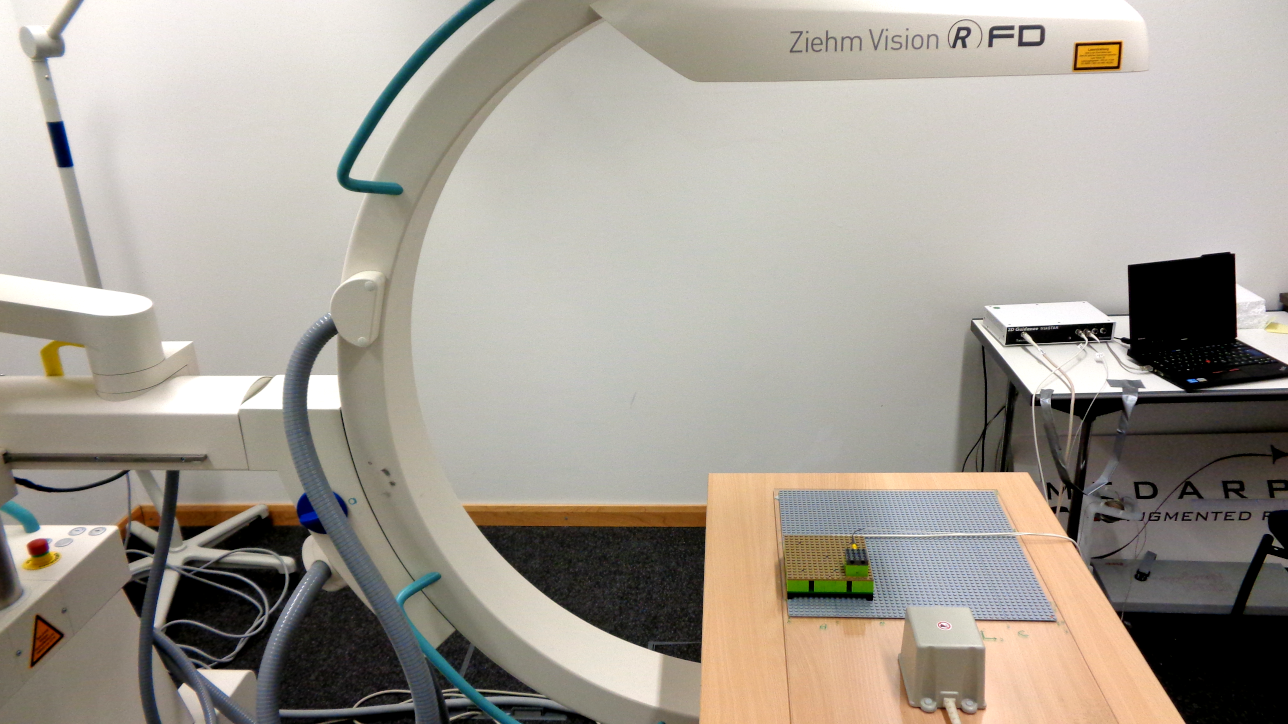}}
		\caption{%
			(a) Calibrated positions (dark dots) on Lego phantom.\\
			(b) Measurement setup in c-arm environment.
		}
	\end{minipage}
	\hspace{0.01\linewidth}
	\captionsetup[subfigure]{labelformat=empty}
	\begin{minipage}{0.5\linewidth}
		\centering
		\subfloat{\includegraphics[height=1.1\linewidth]{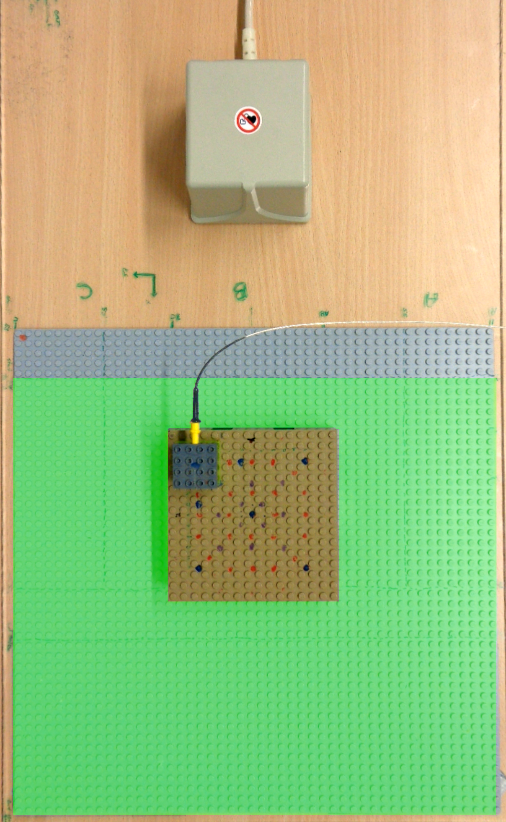}}
		\caption{%
			Lego phantom (brown) on base board.
			Green area marks specified region of the trakSTAR.
		}
		\label{fig:setup}
	\end{minipage}
\end{figure}

In this paper, we present an active \textit{online} error compensation approach for \ac{EMT} navigation in \ac{AAAR}.
First, we collect positional \ac{EMT} data in one laboratory and several \ac{OR} scenarios with different degrees of distortion.
We capture \ac{EMT} sensor positions on a calibrated Lego phantom (\figref{fig:phantom} and 2).
Metallic distortion is artificially introduced to the magnetic field by positioning a c-arm fluoroscopy unit (\figref{fig:carm_near}) in varying alignments next to the Lego setup.
We then use an \ac{ANN} for approximating a function that maps erroneous positions to compensated positions.
The \ac{ANN} is evaluated in an \textit{online} setup to compensate distortions that are not available in the training phase.
As model predictions can be uncertain in regions where training data is sparse, we assess spatial uncertainty inherent to the \ac{ANN}.
This uncertainty evaluation is performed at different regions of the navigation volume with varying availability of training data.
In a final experiment, we simulate a trajectory resembling the guidewire path through the abdominal aorta.
We use our knowledge about \ac{ANN} inherent uncertainty for finding optimal points to acquire x-ray images.
For guidewire insertion in the real \ac{OR}, these x-ray images can be used to rectify uncertain (and hence erroneous) \ac{EMT} sensor positions.
Our simulation provides initial understanding to the trade-off in loss of precision versus reduction in radiation exposure using hybrid tracking.

This is the first work describing an uncertainty-aware online error compensation approach for \ac{EMT} in endovascular surgery.
Our contributions are twofold:
First, we describe a neural network to approximate a spatial compensation function based on relative distances.
The approximation works online in scenarios with unknown distortions.
Second, we assess spatial model-inherent uncertainty of the neural network regression and its effect on positional error.
This analysis provides us insight about the linear relation between model uncertainty and tracking accuracy, and the potential radiation-error trade-off for hybrid guidewire navigation in \ac{AAAR}.

\section{Related Work}
% !TeX spellcheck = en_US

As a comprehensive description of all the active \ac{EMT} error compensation techniques is beyond the scope of the paper, we point the reader to Franz et al. \cite{franz2014} and Kindratenko et al. \cite{kindratenko2000survey} who provide a comprehensive review of this topic.
Instead, in this paper, we mainly focus on the compensation techniques similar to ours.
%Techniques such as lookup-tables [8], interpolation [12] or polynomial regression [5] are proposed, but only work on known distortion characteristics (offline compensation).
Kindratenko et al. \cite{kindratenko2005neural} propose a two hidden layer neural network that outperforms polynomial fits and lookup-table compensation in an offline compensation setup.
%Nearly all of the proposed error compensation algorithms only work in offline setups.
%Online error correction algorithms are conditioned on data from the same distortion scenario in which they will be applied.
%These approaches are hardly applicable in the clinical environment \cite{franz2014}, hence they were barely investigated since the early 2000s.

\textit{Online} compensation approaches use data from additional sensors \cite{sadjadi2016slam} or sensor arrays \cite{ramachandran2016distortion} to map metallic distortions in the tracking volume.
Sadjadi et al. propose a \ac{SLAM} approach that reduces positional error by 67\%, but requires auxiliary sensors to be rigidly attached to the tracked instrument -- which is not applicable to guidewires or catheters in endovascular navigation.

In endovascular surgery, the use of \ac{EMT} is evaluated in several phantom \cite{wood2005, delambert2012}, swine \cite{wood2005, manstad2011} and patient studies \cite{manstad2012}.
These studies show that there is potential for \ac{EMT} to be applied in \ac{AAAR}, with positional errors of up to \SI{5}{\milli\meter}.
%Although maximum positional errors of \SI{5}{\milli\meter} are generally believed to be acceptable, we see potential for improvement to enhance navigation accuracy.
%Improving the reliability of \ac{EMT} makes less x-ray images necessary for recalibration and thus reduces radiation dose.

Neural networks, such as those we use for error compensation in this paper, are black-boxes due to their complexity and non-linearity.
We therefore employ means to make model predictions traceable.
Gal et al. \cite{gal2016dropout} propose to use dropout masks for hidden layers at both training and inference time to obtain a Bayesian approximation for prediction uncertainty in classification problems.
In this paper, we generalize this approximation to learn about the limits of the presented regression approach for spatial error compensation.

\section{Materials}
% !TeX spellcheck = en_US
Positional tracking experiments are performed with an Ascension trakSTAR 3D Guidance system (Northern Digital Inc.) under the use of a \SI{1.8}{mm} sensor.
Positional \ac{EMT} measurement data is collected on a calibrated Lego measurement phantom (repeatability \SI{20}{\micro\meter}) similar to the one proposed by us earlier \cite{kuegler2019}.
\ac{EMT} measurements are performed in laboratory and near a Ziehm Vision 3D c-arm fluoroscopy unit.
Software for interfacing the trakSTAR system is developed in C++.
Compensation models are implemented in Python (Python Software Foundation) using Keras \cite{keras} and tensorflow backend \cite{tensorflow}.
%Nine phantom positions on the planar base board are measured in the specified tracking volume.
%Each position is measured five times per scenario.
%Additionally, measurements are repeated with the phantom rotated around \ang{180} on all nine positions.
%Displacements are only calculated per phantom position on base board, hindering phantom repeatability to contribute to overall measurement uncertainty.

%\begin{table}
%	\begin{tabular}{c|c|c|c|c}
%		scenario & \thead{\#displacements\\before/after\\preprocessing} & \thead{approx. distance\\to c-arm [cm]} & \thead{displacement\\RMSE [mm]} & \thead{max. displacement\\error [mm]} \\ 
%		\hline
%		laboratory & 4950/2877 & - & 0.3854 & 1.1217 \\
%		c-arm-1 & 4950/2065 & 42 & 0.6813 & 1.9963 \\
%		c-arm-2 & 4950/2571 & 23 & 1.0111 & 2.3082 \\
%		c-arm-3 & 4950/2776 & 7  & \textbf{1.6664} & \textbf{5.6468} \\
%		c-arm-4 & 4950/3186 & 36 & 0.5136 & 1.7050 \\
%		\hline
%		c-arm-5 & 1215/832  & 32 & 0.9830 & 3.0729 \\
%		c-arm-6 & 1215/901 & 34 & 0.9364 & 2.6895
%	\end{tabular}
%	\caption{%
%		Datasets collected in different environments.
%		Number of displacements (before and after eliminating redundant distances), distance from c-arm, RMSE and max. displacement error are noted for each dataset.
%		Distances are measured from x-ray source to base board center.
%		Scenarios \carm{5} and 6 are designated for evaluation only.
%		%The c-arm gantry is rotated around 30 degrees in \carm{6}.
%	}
%	\label{tab:scenarios}
%\end{table}

\begin{table}
	\centering
	\begin{tabular}{c|c|c|c|c}
		& scenario & \thead{\#displacements} & \thead{displacement\\RMSE [mm]} & \thead{max. displacement\\error [mm]} \\ 
		\hline
		\multirow{4}{*}{training} & c-arm \SI{7}{\centi\meter} & 870 & 1.386 & 3.586 \\
		& c-arm \SI{8}{\centi\meter} & 870 & 1.292 & 3.239 \\
		& c-arm \SI{9}{\centi\meter} & 870 & 1.192 & 3.221 \\
		& c-arm \SI{10}{\centi\meter} & 870 & 1.101 & 2.994 \\
		\hline
		\multirow{1}{*}{validation} & \shortstack[l]{\\laboratory} & \shortstack[l]{\\870} & \shortstack[l]{\\0.367} & \shortstack[l]{\\0.916} \\
		\hline
		\multirow{3}{*}{evaluation} & c-arm \SI{11}{\centi\meter} & 870 & 1.064 & 2.926 \\
		& c-arm \SI{12}{\centi\meter} & 870 & 1.025 & 1.403 \\
		& c-arm\textsuperscript{1} \SI{30}{\centi\meter} & 870 & 0.743 & 1.671 \\
		& c-arm\textsuperscript{2} \SI{50}{\centi\meter}\ & 870 & 0.639 & 1.403 \\
	\end{tabular}
	\caption{%
		Datasets collected in varying distances to c-arm and in a laboratory setup.
		Number of displacements, RMSE and max. displacement error are noted for each dataset.
		Distances to c-arm are measured from x-ray source to base board center.
		c-arm\textsuperscript{1}:  gantry rotated at $30^\circ$, c-arm\textsuperscript{2}: gantry rotated at $60^\circ$
	}
	\label{tab:scenarios}
\end{table}

\section{Methods}
% !TeX spellcheck = en_US
First, training and evaluation datasets are acquired in one laboratory and multiple c-arm scenarios.
We describe the acquisition and preprocessing in \secref{sec:dataacquisition}.
Acquired datasets are then used for training neural networks for \ac{EMT} error compensation, which we describe in \secref{sec:models}.
We use our \ac{ANN} in four experimental setups.
In the first experiment, the \ac{ANN} is trained on a multitude of datasets for \textit{online} compensation (\secref{sec:acrossscenarios}).
Afterwards, we perform an \textit{offline} evaluation to compare the \acp{ANN} to a similar compensation approach (\secref{sec:scenariowise}).
We then evaluate spatial model uncertainty for the online model in \secref{sec:epistemic}.
Finally, in a simulation experiment (\secref{sec:simulation}), we use model uncertainty to find a threshold for recalibration.

\subsection{Data acquisition and preprocessing}
\label{sec:dataacquisition}
% more info on #samples in kuegler2019 and rebuttal letter
Data points are captured by sequentially positioning a Lego block with an embedded \ac{EMT} sensor on ten calibrated positions (see \figref{fig:phantom}) of the phantom.
Random \ac{EMT} error is eliminated by taking the median of 500 samples.
%\TODO{change: Duplicate measurements are removed from datasets to speed up training and to prevent overfitting redundant data.}
%\TODO{change: Due to the time consuming and tedious data acquisition process, we only measure positions in a two dimensional plane.
%Recording a single scenario (five measurements per board position) takes two hours on average.
%As manual data acquisition effort rises exponentially with each additional \ac{DOF}, we therefore restrict to two of the six \ac{DOF} that \ac{EMT} sensors usually provide.}
%We expect that considering more \ac{DOF} (by measurements in different heights and orientations) improves overall compensation performance and reduces epistemic uncertainty.
We collect positional datasets in multiple scenarios with artificial distortion and in a laboratory scenario.
Each distorted scenario uses a different c-arm alignment with respect to the Lego phantom.
\tabref{tab:scenarios} shows the datasets collected for the experiments in \secref{sec:acrossscenarios} and \secref{sec:scenariowise}.
%Errors are calculated
Positional data are collected in three phantom elevations in steps of \SI{9.6}{\milli\meter} (height of one Lego brick).
With each c-arm position, we also measure positions with the phantom rotated by $180^\circ$ around its azimuth axis.
Error values are calculated as $e = ||x_2 - x_1|| - y$, where e is error, $x_1$ and $x_2$ are two different measuring points and $y$ is the respective ground truth distance on the Lego board.

\subsection{Error compensating neural networks}
\label{sec:models}
We mitigate systematic positional error by approximating a compensation function that maps erroneous to compensated points.
The compensation function is approximated by a three-layer \ac{ANN} with 32 units per layer.
These parameters as well as the batch size (512) are estimated by grid search.
The \ac{ANN} uses leaky ReLU activations ($\alpha=0.01$) in the hidden layers to prevent vanishing gradients.
%As the training procedures differ between use case scenarios (\textit{unseen} and \textit{known} distortions), they are further described in sections \ref{sec:acrossscenarios} and \ref{sec:scenariowise} respectively.

%The compensation architecture consists of two symmetric paths for the input point pairs.
%\TODO{rewrite: Both paths contain modules with shared weights to ensure that the compensation function can be used for absolute point correction after training.}
The compensation function has four input units for x, y, z and the trakSTAR quality indicator, which are all normalized to an interval $[0, 1]$ to improve model stability.
This normalization is reverted after the final layer, which contains three units with linear activations for x, y and z coordinates.
%The normalization is reverted before euclidean distance calculation, so that resulting distances can be compared to ground truth distances.

\begin{figure}
	\centering
	\includegraphics[width=0.7\linewidth]{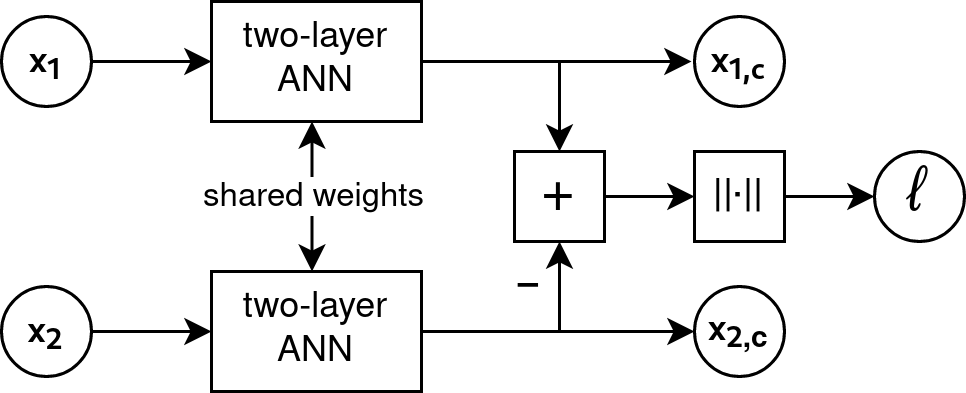}
	\caption{%
		Neural network model for point compensation.
		$x_1$, $x_2$ are input points (x, y, z, quality), $x_{1,c}$ and $x_{2,c}$ are compensated output points.
		$\ell$ is the displacement distance we use for computing the MSE-loss.
	}
	\label{fig:nnarchitecture}
\end{figure}

During training, two \acp{ANN} with shared weights are arranged in parallel, as illustrated in \figref{fig:nnarchitecture}.
We can therefore train the compensation function on relative displacements, but use the trained function for absolute point predictions.
Training on relative displacements between positions ensures that the exact distance of phantom to the field generator center does not need to be measured \cite{kuegler2019}.
Hence, this approach circumvents the need for a second measurement standard to capture absolute positions, which would contribute to overall measurement uncertainty.
In the training phase, the displacement error (mean-squared-error) is minimized by Adam optimizer \cite{kingma2014adam} (learning rate = $0.01$):

\begin{equation}\label{eq:loss}
\mathcal{L} = ||f(x_{2},q_{2},\omega) - f(x_{1},q_{1},\omega)||_2 - y
\end{equation}

\noindent where $f$ denotes the learned compensation function approximated by the \ac{ANN}, $y$ is the ground truth displacement length, $x_{1}$, $x_{2}$ are measured \ac{EMT} points, $q_{1}$ and $q_{2}$ are respective quality indicator values and $\omega$ is the matrix of learned weights.

As mentioned earlier, we add a quality indicator value that is reported by the trakSTAR system along with every measurement, as additional input to the compensation models in expectation of better generalization performance across scenarios.
According to the trakSTAR user manual \cite{manual3dg}, the quality value is computed from an internal error indication $\epsilon$ and four user-defined quality-parameters:

\begin{equation}\label{eq:quality}
Q = S \cdot (\epsilon - (b + m \cdot r))
\end{equation}

\noindent where $S$, $b$, and $m$ denote user parameters (sensitivity, offset, slope), $r$ is the sensor-transmitter range.
We obtain \textit{raw} quality values by setting the user parameters to $S=1,~b=0,~m=0$.

For comparison, we implement mixed-term polynomial regression models as proposed by Kügler et al. \cite{kuegler2019}.
Both compensation models are trained on pairs of \ac{EMT} sensor positions and corresponding ground truth distances on the Lego phantom.

\subsection{Compensation of unseen distortions}
\label{sec:acrossscenarios}

We train the \ac{ANN} on data from four c-arm scenarios (see \tabref{tab:scenarios}).
For model validation, we use the data obtained in the laboratory setup.
The trained model is evaluated on the remaining four datasets (\figref{fig:interscenarioresults}).
Likewise, we train and evaluate the polynomial regression model for comparison to the compensation approach proposed by Kügler et al. \cite{kuegler2019}.

\subsection{Known distortion compensation}
\label{sec:scenariowise}

In this experiment, we evaluate \ac{ANN} models in the same c-arm setup in which they are trained (\textit{offline} compensation).
This experiment measures the best-case outcome for learning based compensation.
We examine compensation for all scenarios in \tabref{tab:scenarios} individually, where training and evaluation sets are chosen to be spatially independent.
The data is divided into training/validation/testing sets with a split of 45/5/50.

\subsection{Model uncertainty evaluation}
\label{sec:epistemic}
Model-inherent uncertainty for the \ac{ANN} is estimated by applying $10\%$ dropout during training and at inference time (Monte Carlo Dropout \cite{gal2016dropout}).
We take 3000 samples from the distribution of compensated output positions to obtain a Bayesian approximation of model-inherent uncertainty.
Spatial uncertainty is expressed as the standard deviation \small{$\sigma = \sqrt{\sigma_x^2 + \sigma_y^2}$}~\normalsize for each point $(x, y)$ in the planar full-base-board dataset.
Distributions of model predictions cannot be assumed to be Gaussian (see section 4.1 in \cite{kuegler2019}), so that the 68-95-99.7 rule for confidence interval approximation does not apply here.

To examine spatial uncertainty for a large portion of the specified tracking volume, we collect a dataset in the 2D plane by moving the phantom to different positions on the gray Lego board.
This dataset contains 10598 different displacements, collected in six different alignments of the c-arm to the tracker.
We use this dataset for training a neural network analogously to \secref{sec:acrossscenarios}, but with modifications to the \ac{ANN} layout.
That is, a neural network with two layers, 64 units per layer and two output neurons (x, y) is employed for the following evaluations.

In addition to the training set, we collect measurement data for evaluation from all Lego points within the specified area (green area in \figref{fig:setup}).
Unlike the training sets, this dataset also contains phantom points that were not calibrated.
As phantom uncertainty ($\approx~\SI{20}{\micro\meter}$ according to \cite{kuegler2019}) is negligible compared to model-inherent uncertainty we want to examine, this simplification is valid.
This measurement allows for an evaluation of spatial epistemic uncertainty over the whole base board (see \figref{fig:heatmapboard}).

%\TODO{change: To examine spatial uncertainty of the trained \ac{ANN}, we collect an additional dataset next to the c-arm.
%Unlike the datasets shown in \tabref{tab:scenarios}, positions in this dataset cover the \textit{whole} specified region (blue frame in \figref{fig:setup}) of the trakSTAR system.
%Moreover, we position the sensor on all phantom positions and do not restrict to calibrated positions only.
%As phantom uncertainty ($\approx~\SI{20}{\micro\meter}$ according to \cite{kuegler2019}) is negligible compared to model-inherent uncertainty, this simplification is valid.
%We evaluate spatial epistemic uncertainty for the whole base board dataset (see \figref{fig:heatmapboard}).}

\begin{figure}[h]
	\centering
	\begin{minipage}[t]{0.48\linewidth}
		\subfloat{\includegraphics[width=0.36\linewidth]{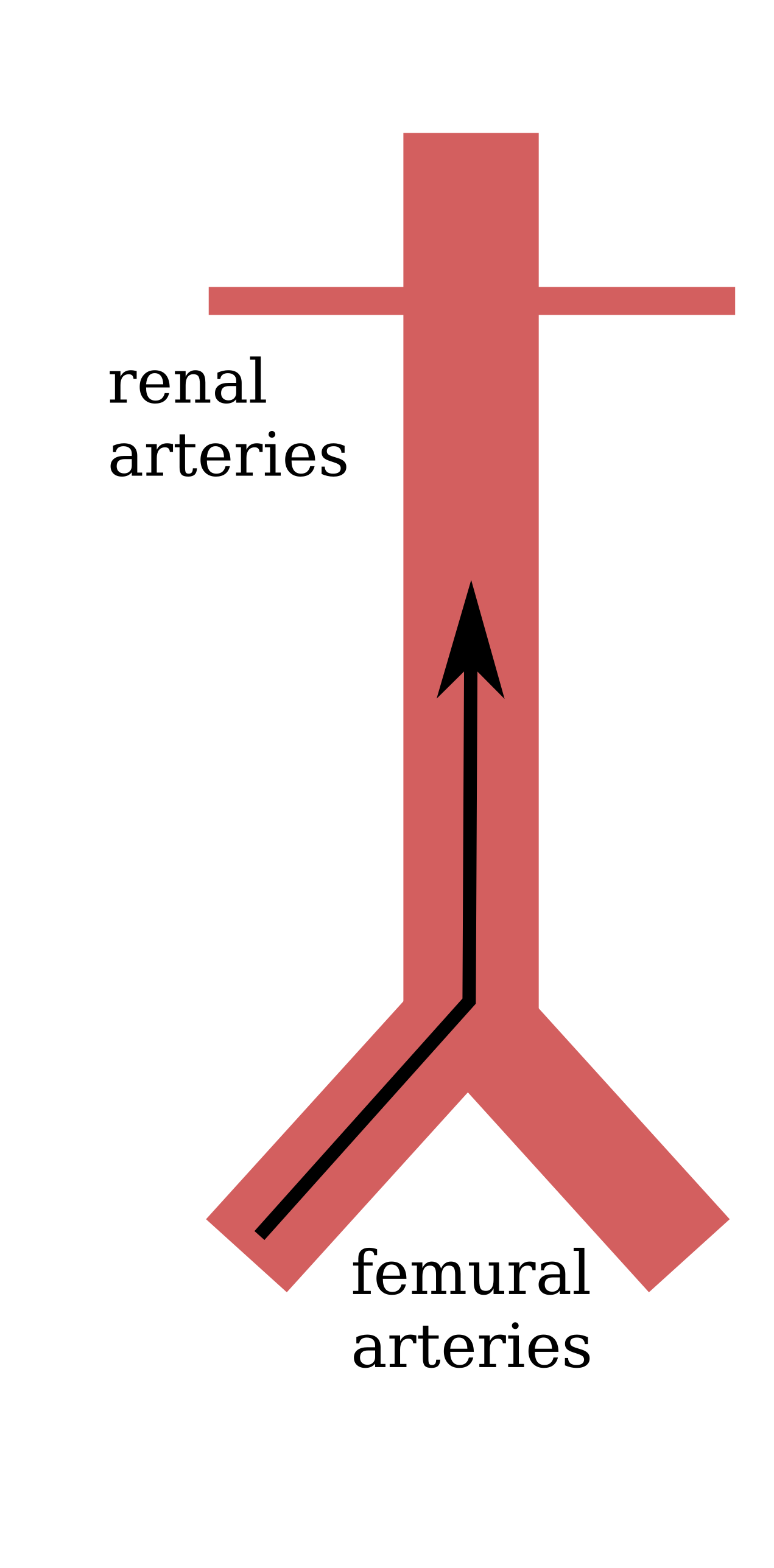}}
		\subfloat{\includegraphics[width=0.64\linewidth]{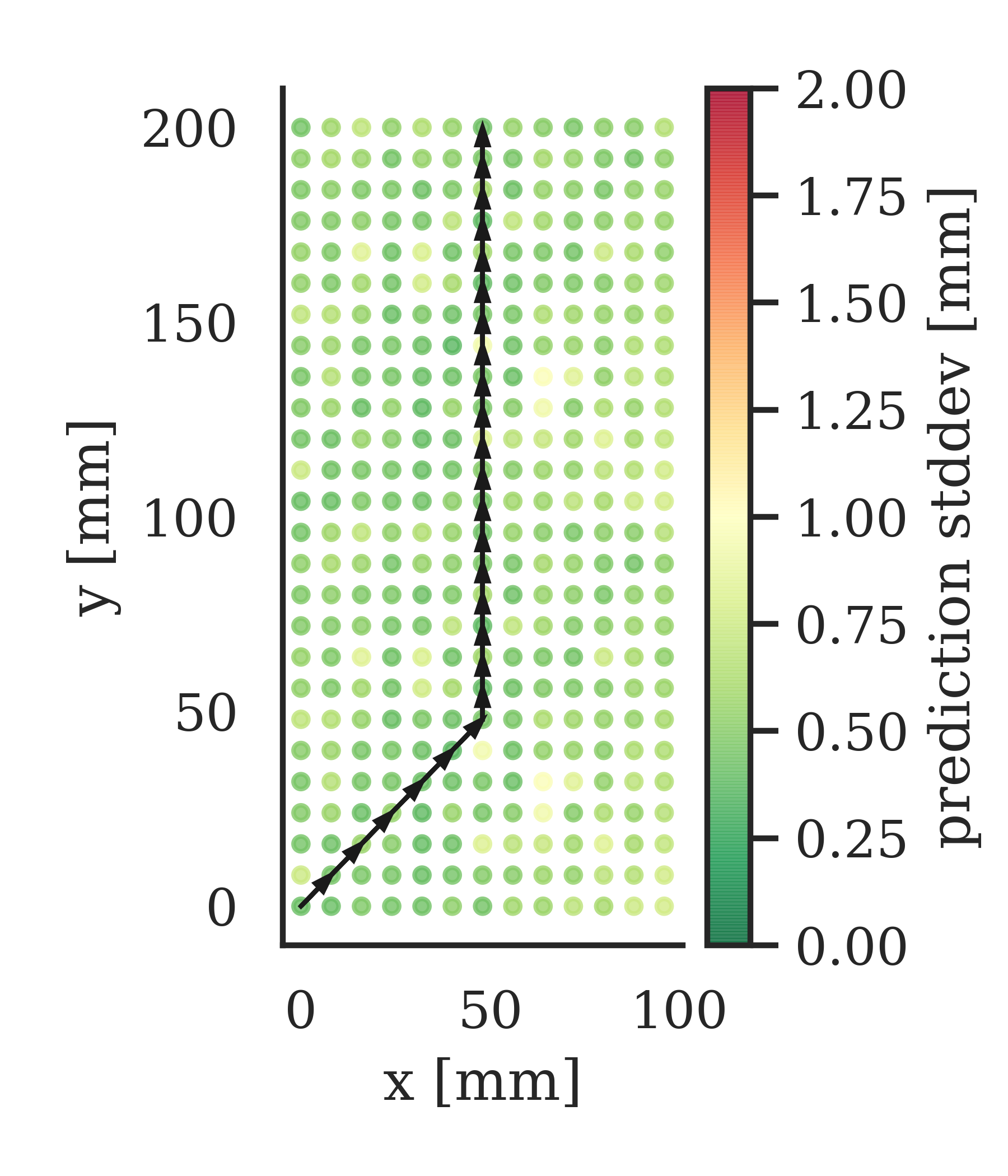}}
		\caption{%
			Schematic abdominal aortic anatomy with guide wire path (left).
			Virtual EMT sensor trajectory segments (black arrows) in center of specified tracking volume (right).
			Dots correspond to measurement positions on Lego phantom.
		}
		\label{fig:simboard}
	\end{minipage}
	\hspace{0.01\linewidth}
	\begin{minipage}[t]{0.49\linewidth}
		\subfloat{\includegraphics[width=1\linewidth]{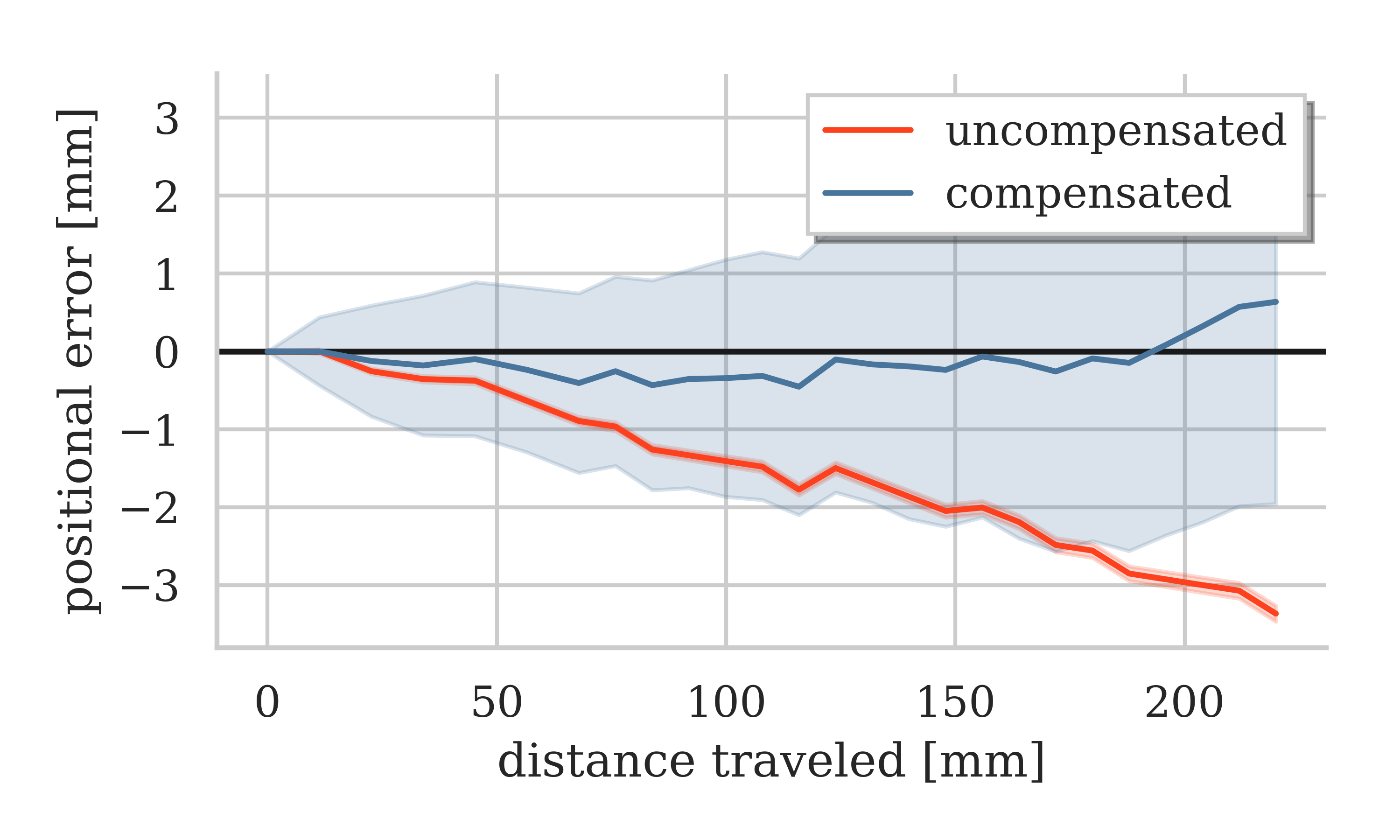}}
		\caption{%
			Error development along simulated trajectories with (blue) and without (red) compensation.
			Filled area depicts uncertainty.
		}
		\label{fig:errordevelopment}
	\end{minipage}
\end{figure}

Although the symmetric \ac{ANN} is trained and validated on pairs of positions and respective ground truth distances, a single trained \ac{ANN} can be used for absolute point compensation during inference (compare $x_{1,c}$ and $x_{2,c}$ in \figref{fig:nnarchitecture}).
In this experiment, we let our trained \ac{ANN} predict compensated positions for the whole baseboard.
Absolute positional error is then estimated by calculating the measured distances to adjacent points in a Moore neighborhood ($r = 3$) \cite{mooreneighborhood} and averaging the error.

\subsection{Simulated hybrid AAAR intervention}
\label{sec:simulation}

Based on real \ac{EMT} data, we simulate a sensor moving on a path inside a virtual abdominal aorta.
This simulation is inspired by guidewire insertion in \ac{AAAR} using hybrid navigation.
Path shapes are motivated by those of abdominal aortae, as can be seen in \figref{fig:simboard}.
Since the average abdominal aorta is \SI{20}{\centi\meter} to \SI{25}{\centi\meter} in length, a simulated path of \SI{21.9}{\centi\meter} is chosen.

Start and end points of each path segment are taken from the full-base-board dataset.
Comparing the distances between segment start and end points to respective ground truth distances yields positional error.
Uncertainty is determined position-wise as described in \ref{sec:epistemic}.

In hybrid navigation, guidewire position can be precisely recalibrated by x-ray pose estimation \cite{kugler2018i3posnet} and fiducial registration \cite{wood2005, manstad2011}.
Correcting the \ac{EMT} sensor location by an x-ray image exposes the patient to radiation, so that recalibrations should rarely be employed.
Hence, we are facing a trade-off between radiation dose and tracking accuracy.
We introduce the concept of recalibration to our simulation by resetting error and accumulated uncertainty at calculated recalibration points.

We evaluate two different strategies for determining when to perform the recalibration in our simulation:
A) We choose recalibration points based on model uncertainty.
Recalibration is performed when accumulated model uncertainty exceeds a certain threshold $\tau$.\\ 
B) We simulate recalibration in defined constant intervals based on traveled distance.
We simulate the recalibration process for different adaptive thresholds $\tau$.
The same is done for different uniform distance intervals ranging from \SI{0}{\centi\meter} to \SI{21.9}{\centi\meter}.

\section{Results}
% !TeX spellcheck = en_US

\begin{figure}[t]
	\centering
	\begin{minipage}{1\linewidth}
		\centering
		\subfloat{\includegraphics[width=0.6\linewidth]{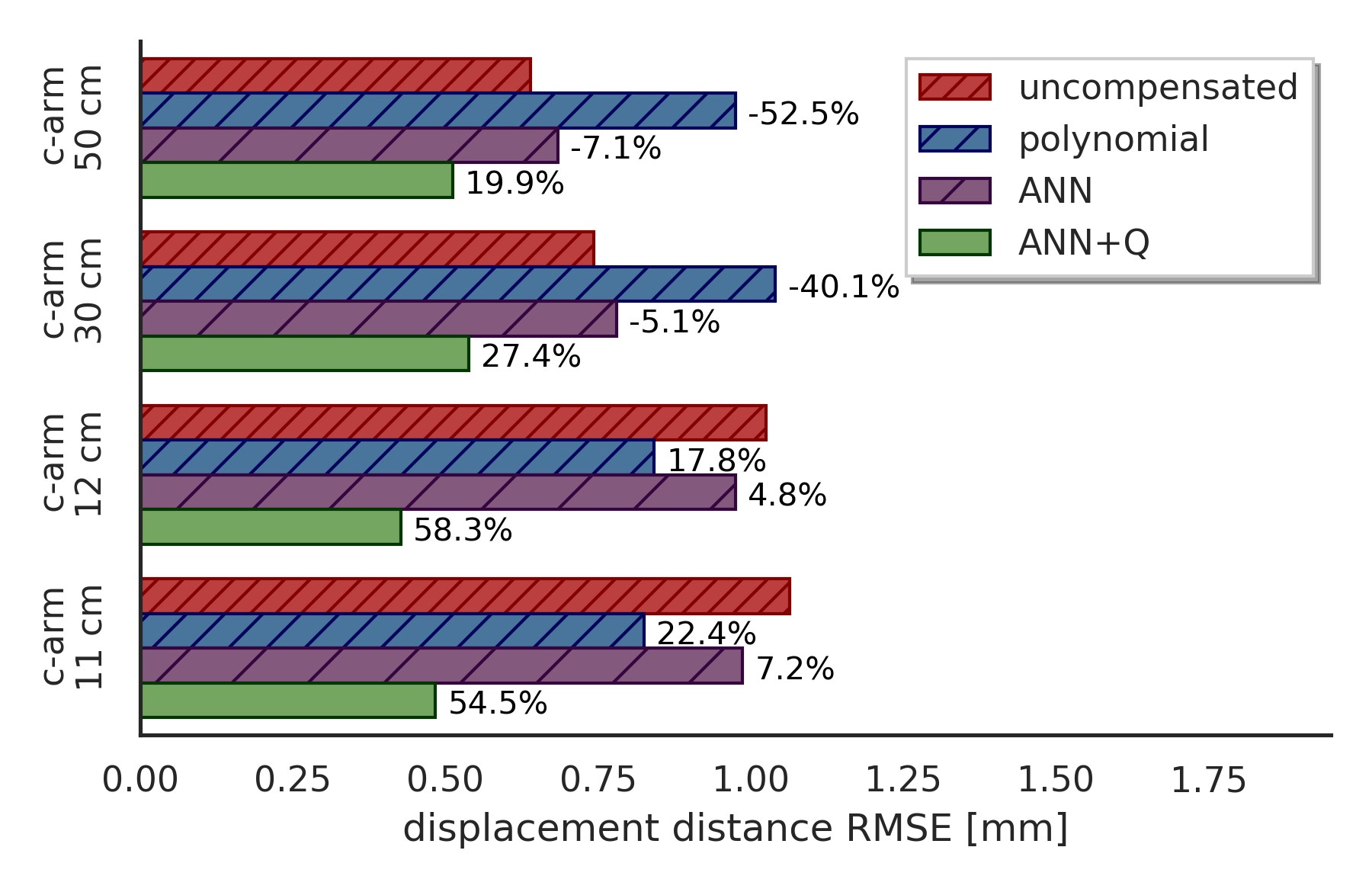}}
		\subfloat{\includegraphics[width=0.4\linewidth]{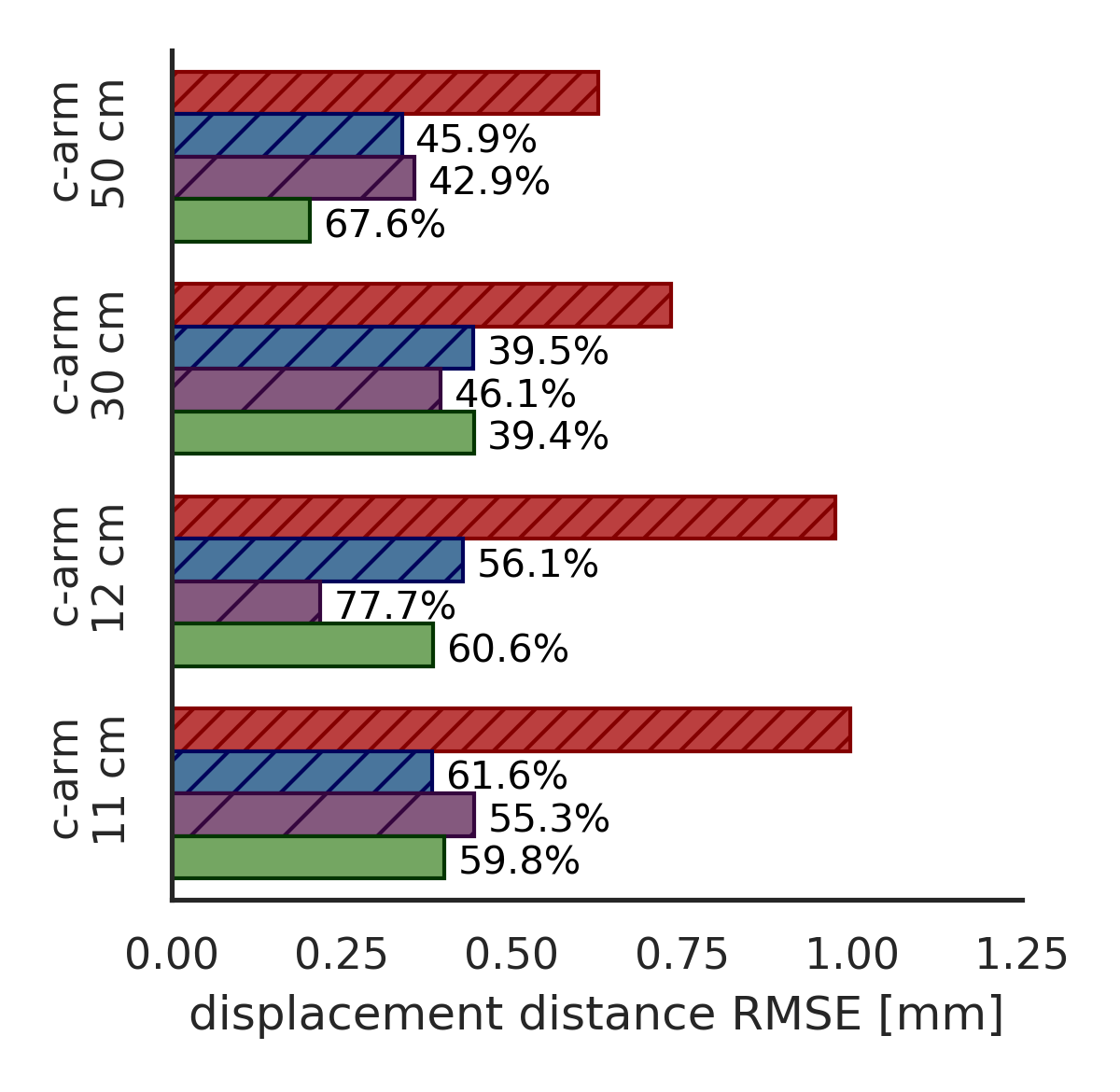}}
		
	\end{minipage}
	\caption{%
		Comparison of compensation performance in scenarios with \textit{unseen} (left) and \textit{known} (right) distortions.
		ANNs with and without quality indicator (Q) are compared to polynomials.
		Percentages denote error reduction.
	}
	\label{fig:interscenarioresults}
\end{figure}

\paragraph{Error compensation}
In unseen scenarios (\secref{sec:acrossscenarios}), \acp{ANN} clearly outperform polynomial regression models (\figref{fig:interscenarioresults}).
However, the compensated \ac{EMT} error does not reach the results achieved by offline compensation.
We observe that including the quality indicator in the model input improves generalization abilities of the neural network.
Scenario-wise compensation (\secref{sec:scenariowise}) appears to be a simple task for both polynomial fits and \acp{ANN}, as we can compensate between $35\%$ and $75\%$ of error in each scenario.
Both algorithms can reduce tracking error to sub-millimeter values in this offline setup.

\begin{figure}[h]
	\centering
	\begin{minipage}[t]{0.53\linewidth}
		\centering
		\subfloat{\includegraphics[width=1.0\linewidth]{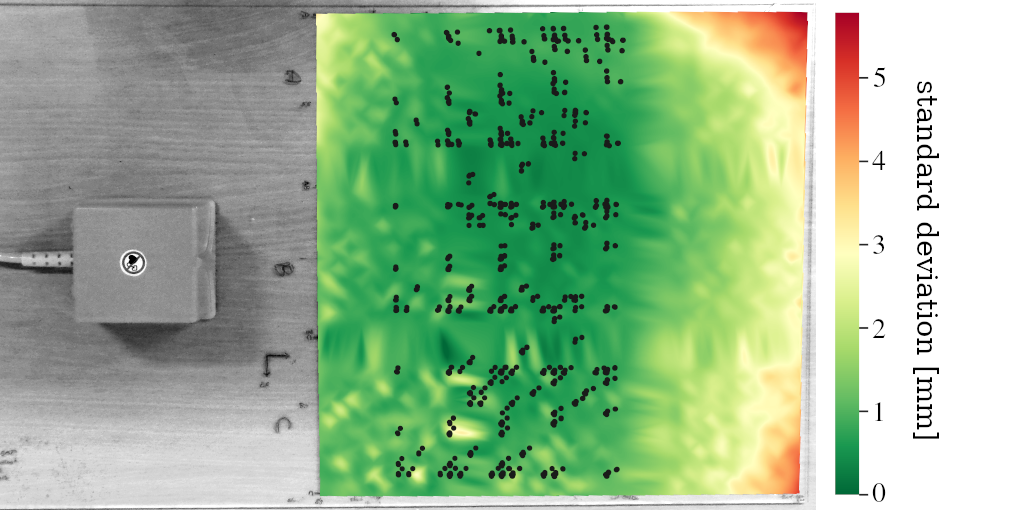}}
		\caption{%
			Model-inherent uncertainty map projected onto the Lego base board.
			Dark spots mark measurement points used for training the ANN.
		}
		\label{fig:heatmapboard}
	\end{minipage}
	\hspace{0.01\linewidth}
	\begin{minipage}[t]{0.43\linewidth}
		\subfloat{\includegraphics[width=1.0\linewidth]{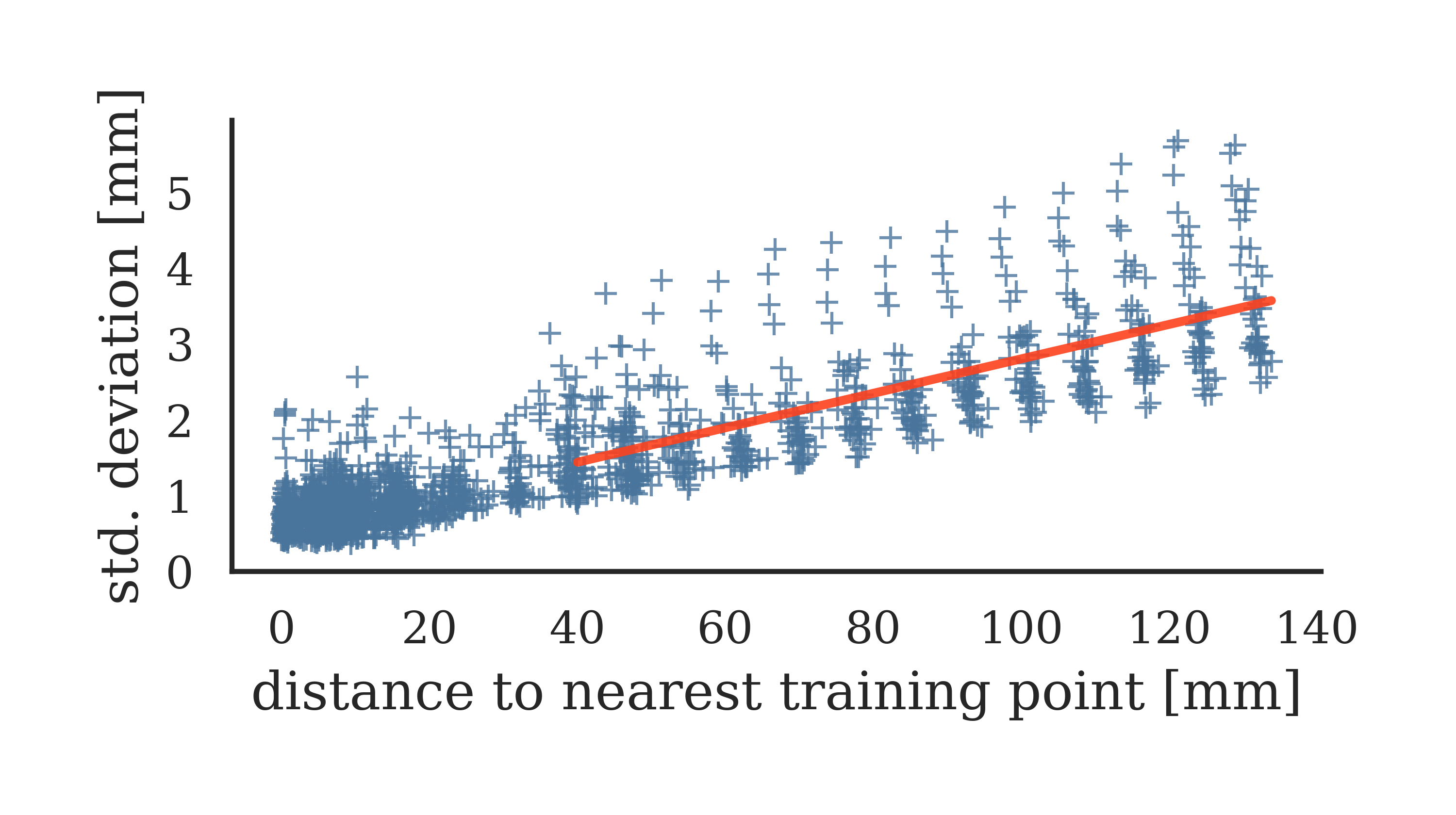}}
		\caption{%
			Relationship of \ac{ANN} uncertainty to distance to nearest training point.
			Red line shows linear regression.
		}
		\label{fig:stddensity}
	\end{minipage}
\end{figure}

\begin{figure}[h]
	\centering
	\captionsetup[subfigure]{labelformat=empty}
		\begin{minipage}[t]{0.48\linewidth}
		\subfloat{\includegraphics[width=1.0\linewidth]{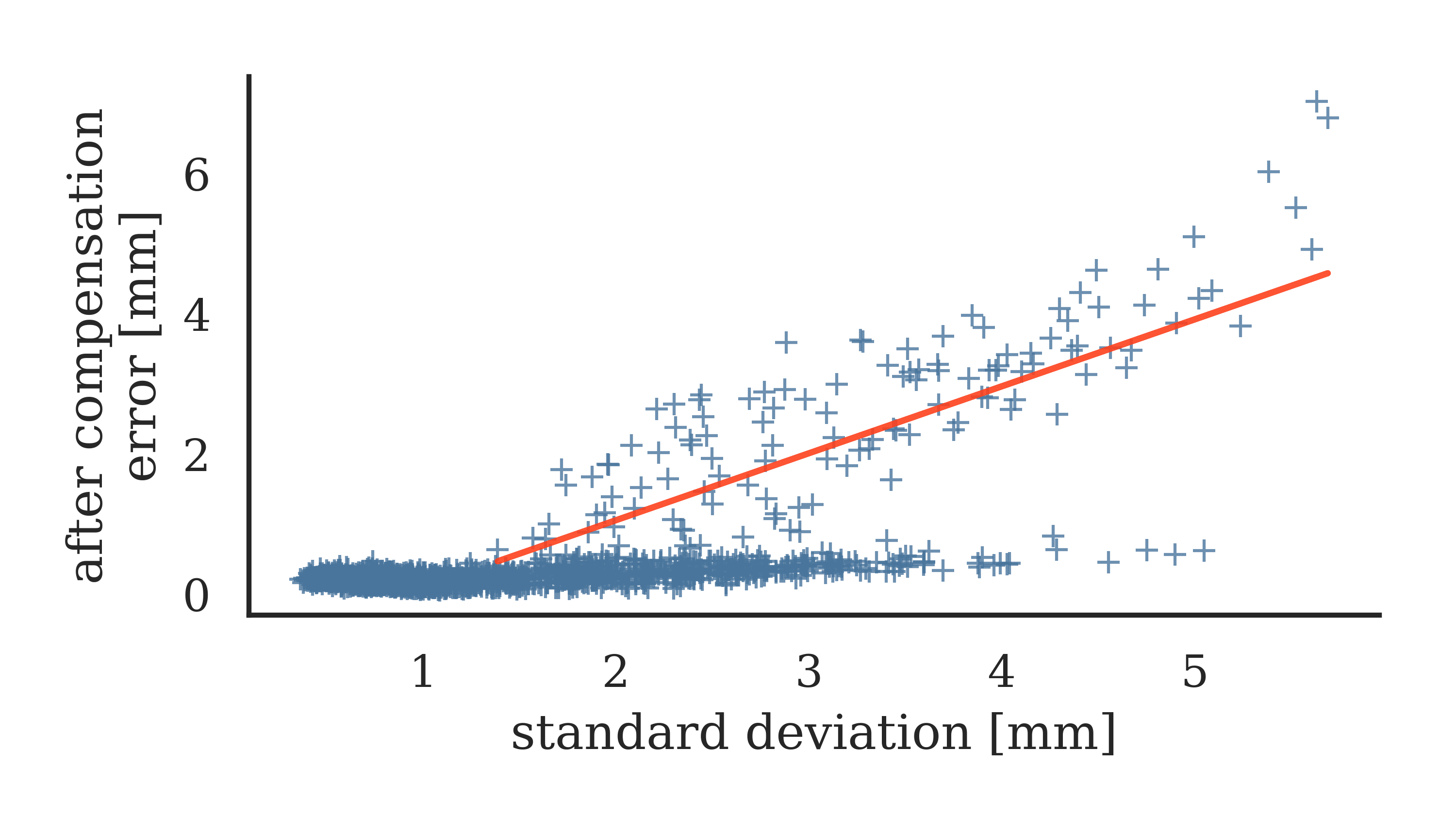}}
		\caption{%
			Relationship between \ac{ANN} uncertainty and compensation error.
			Red line shows linear regression.
		}
		\label{fig:stderr}
	\end{minipage}
	\hspace{0.01\linewidth}
	\begin{minipage}[t]{0.48\linewidth}
		\subfloat{\includegraphics[width=1.0\linewidth]{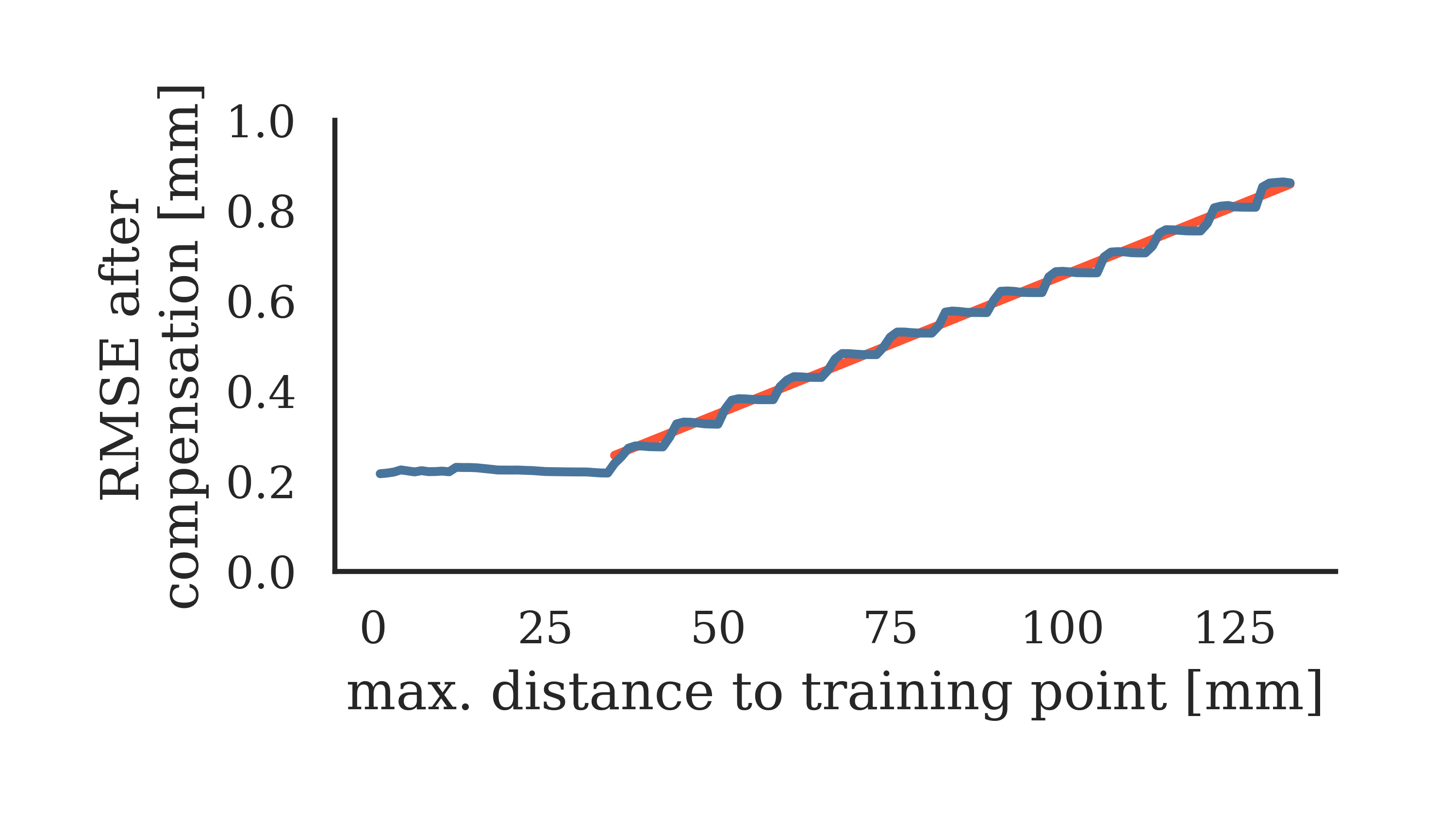}}
		\caption{%
		Relationship between max. distance to training point and compensation RMSE.
		Red line shows linear regression.
		}
		\label{fig:rmsedensity}
	\end{minipage}
\end{figure}

% model uncertainty
\paragraph{Model uncertainty evaluation}
We compare error after compensation to model uncertainty, finding that both quantities are correlated (Fig.~\ref{fig:stderr}).
Hypothesizing that model-inherent uncertainty grows with less training points nearby, we measure distances between points in the evaluation set and the nearest point in the training set.
The resulting distances serve as a proxy measure for training point density.
Figure \ref{fig:stddensity} shows the relationship between training point density and the resulting model uncertainty.
We observe that for distances to next training point greater than \SI{35}{\milli\meter}, error after compensation grows linearly with decreasing point density (\figref{fig:rmsedensity}).

\begin{figure}[t]
	\centering
	\begin{minipage}{1\linewidth}
		\centering
		\subfloat{\includegraphics[width=0.5\linewidth]{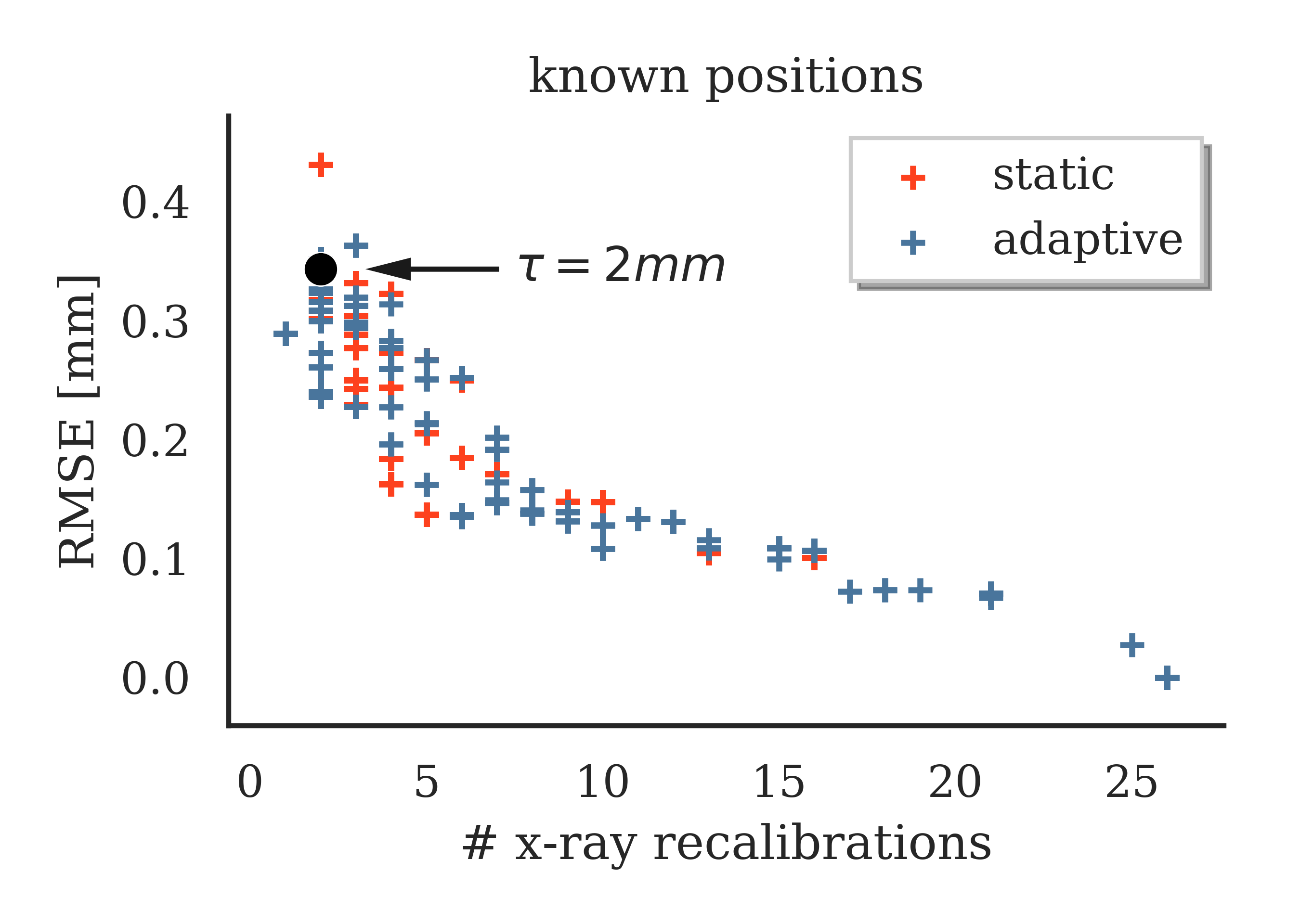}}
		\subfloat{\includegraphics[width=0.5\linewidth]{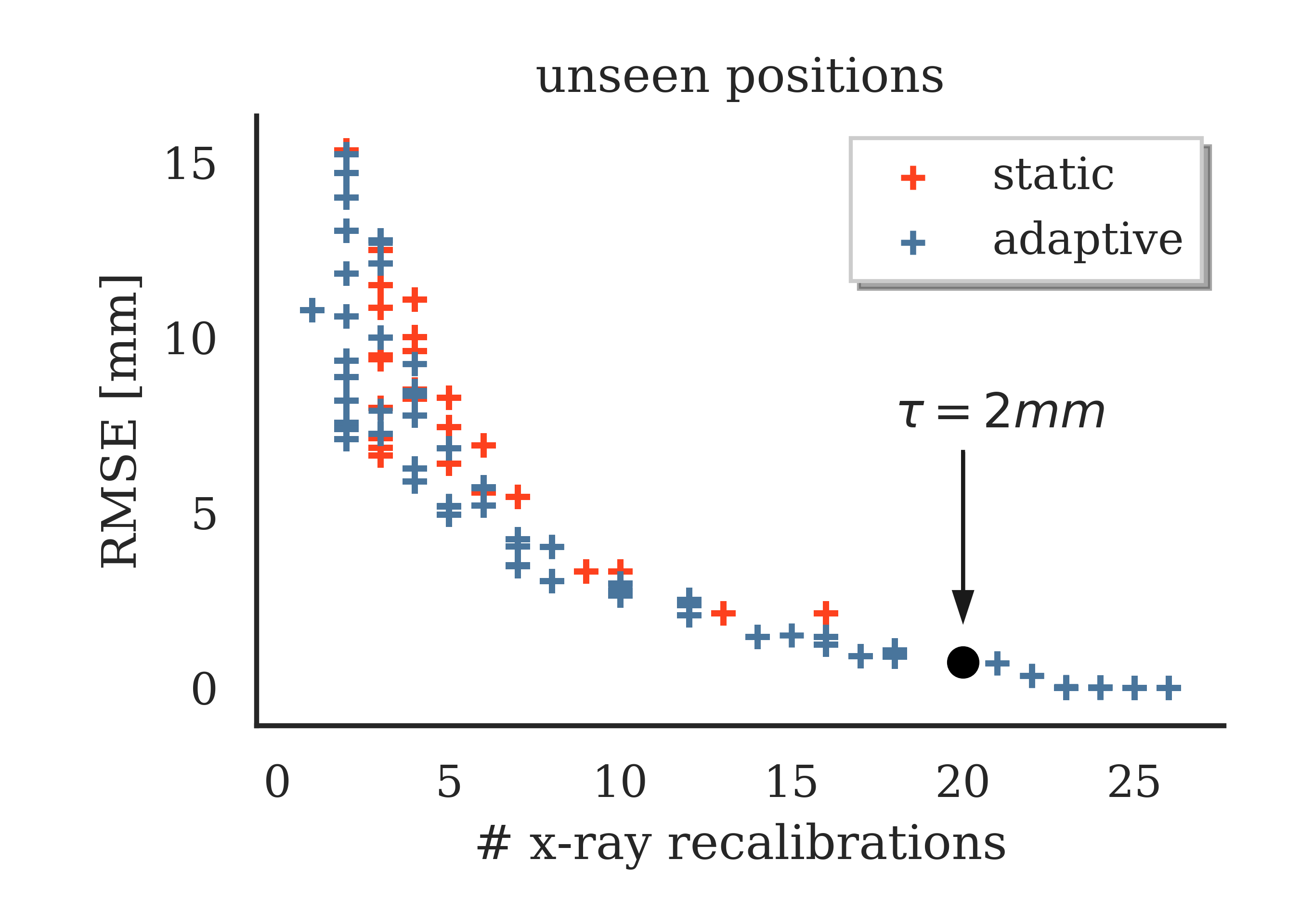}}
	\end{minipage}
	\caption{%
		Pareto front for radiation vs. error trade-off at seen (left) and unseen (right) regions of specified tracking volume.
	}
	\label{fig:simpareto}
\end{figure}

\paragraph{Simulated hybrid AAAR intervention}
Along with the sensor traveling along its path, uncertainty accumulates by \small{$\sigma_{n+1}=\sqrt{\sum_{i=1}^n\sigma_i^2}$}~\normalsize.
Figure \ref{fig:errordevelopment} shows how error and uncertainty develop during virtual guidewire insertion with and without x-ray recalibration.
The trade-off between required x-ray recalibrations and tracking error for uncertainty-based (blue, adaptive) and distance-based (red, static) triggering of x-ray recalibrations is illustrated in Figure \ref{fig:simpareto}.
Choosing a threshold of $\tau=\SI{2}{\milli\meter}$, as motivated by Figure \ref{fig:stderr}, yields a good compromise between tracking error and radiation exposure in both seen and unseen scenarios.

\section{Discussion}
% !TeX spellcheck = en_US
Although localization error of \SI{4}{\milli\meter} are believed to be acceptable in endovascular surgery \cite{wood2005}, improving \ac{EMT} accuracy raises the overall trust of the hybrid navigation system.
In this work, we have shown that \acp{ANN} can improve positional tracking to achieve sub-millimeter accuracy in an online setting.
With higher localization accuracy, less x-ray images are needed for navigation.

Spatial uncertainty can be examined for \ac{ANN} models and it should be utilized as a measure for model validation whenever compensation algorithms are employed.
On the one hand, knowledge about model uncertainty can be used to refine phantom design.
We have found that our \ac{ANN} model requires a training point spacing of \SI{35}{\milli\meter} to be sufficiently accurate, which should be considered in future experimental designs.
On the other hand, our simulation experiments show that knowledge about model uncertainty can be exploited to minimize radiation exposure in the online hybrid setting.
We envision that model-inherent uncertainty assessment will become an essential part of future \ac{EMT} error compensation approaches.

The presented method can be further improved by conducting more realistic evaluations.
Currently, assumptions about radiation reduction are solely made on the basis of simulations.
Consequently, the findings presented in this paper should be assessed in realistic phantom or cadaver studies.

In addition, the rotational \acp{DOF} need to be considered for realistic evaluations.
Especially the roll angle of \ac{EMT} sensors are of great importance in endovascular procedures and will thus be a major subject of our future work.
Our current evaluations show that \ac{ANN} can compensate static error to sub-millimeter values with only a single sensor.
However, the effort of data collection in multiple scenarios with all \ac{DOF} poses a problem that still needs to be solved, for instance by automatized data acquisition.

Furthermore, only one source of metallic distortion (c-arm) is considered in our experiments.
In the real \ac{OR}, other metallic artifacts contribute to measurement error in addition to the c-arm.
Considering additional artifacts, such as the patient bed, and the rotational \ac{DOF} might require more complex models than the proposed neural network.

\section{Conclusion \& Future Work}
% !TeX spellcheck = en_US

%%%% Summary %%%%
In this paper, we present a novel active error compensation framework for \ac{EMT} in endovascular surgery.
We introduce neural networks capable of generalizing across distortion scenarios in single-sensor configuration while providing sub-millimeter accuracy.
We also quantify the positional uncertainty of the error compensating neural network.
When error compensated \ac{EMT} reaches its limits, we show that knowledge about positional uncertainty helps to get \ac{EMT} navigation back on track.
Our work suggests inherent limits of spatial uncertainty that can only be realized when \ac{EMT} and the compensation scheme are evaluated in tandem.
In future phantom evaluation protocols, we will consider these spatial uncertainty limits.

%%% Future Work %%%
In the future, we will work on automatized data acquisition protocols in order to extend our approach to more than two \ac{DOF}.
Moving towards more realistic evaluations, we will evaluate our method in a hybrid setup with 3D printed aortic phantoms and additional metallic artifacts.

\section*{Compliance with Ethical Standards}\label{CES}

\textbf{Disclosure of potential conflicts of Interest: }
This research was partially funded by the German Research Foundation and BiomaTiCS Research Platform UMZ.
The authors declare that they have no conflict of interest.

\noindent\textbf{Research involving Human Participants and/or Animals: }This article does not contain any studies with human participants or animals performed by any of the authors.

\noindent\textbf{Informed consent: }This articles does not contain patient data.

%\begin{acknowledgements}
%\end{acknowledgements}

\section*{Conflict of interest}
The authors declare that they have no conflict of interest.

\bibliographystyle{spbasic}      % basic style, author-year citations
\bibliography{99_ipcai2020}   % name your BibTeX data base

\end{document}